\documentclass[fleqn,twoside,epsf]{article}
\usepackage{espcrc2}

\usepackage{graphicx}

\newcommand{\AmS}{{\protect\the\textfont2
  A\kern-.1667em\lower.5ex\hbox{M}\kern-.125emS}}
 
\newcommand{\be}{\begin{equation}}
\newcommand{\ee}{\end{equation}}
\newcommand{\beq}{\begin{eqnarray}}
\newcommand{\eeq}{\end{eqnarray}}
 
\title{ Momentum dependence of the N to $\Delta$ transition form factors
\thanks{Talk presented by C.~Alexandrou}}

\author{C.~Alexandrou\address[Cyprus]{Department of Physics,
University of Cyprus, CY-1678 Nicosia, Cyprus},
Ph.\ de Forcrand\address{ETH-Z\"urich, CH-8093 Z\"urich and CERN Theory Division, CH-1211 Geneva 23, Switzerland},
H.~Neff\address{Physics Department, Boston 
University, Boston,  Massachusetts 02215, USA},
J.~W.~Negele\address[MIT]{Center for Theoretical Physics, Laboratory for
Nuclear Science and Department of Physics, Massachusetts Institute of
Technology, Cambridge, Massachusetts 02139, USA},
W. Schroers\thanks{Supported by the Alexander 
von Humboldt Foundation}\addressmark[MIT]
and
A.~Tsapalis\thanks{Supported by the Levendis Foundation}\addressmark[Cyprus]
}

\begin{document}
 
\begin{abstract}
We present a new method to  determine the momentum dependence of the N to
$\Delta$ transition form factors and demonstrate its effectiveness
in the quenched theory at $\beta=6.0$ on a $32^3 \times 64$ lattice.
We address a number of technical
issues such as the optimal combination of matrix elements and the simultaneous
overconstrained analysis of all lattice vector momenta contributing to
 a given momentum transfer squared, $Q^2$. 
\vspace{1pc}
\end{abstract}

\maketitle

\vspace*{-1.7cm}

\section{Introduction}
The N to $\Delta$ transition form factors encode important information
on hadron deformation and have been studied carefully  in recent experiments~\cite{experiment}. In this
work we present the  first lattice evaluation of the momentum dependence of
the magnetic dipole, M1,   the
electric quadrupole, E2, and the Coulomb quadrupole, C2, transition amplitudes. They are calculated
 in the quenched approximation on a lattice of size
$32^3\times 64$ at $\beta=6.0$ with Wilson fermions 
with sufficient
accuracy  to exclude a zero value of E2 and C2 at low $Q^2$.
This accuracy  is achieved by
applying   two novel methods: 
 1) We use  an interpolating field  for  the
$\Delta$ that allows a maximum number of statistically 
distinct lattice measurements 
 contributing to a given $Q^2$.
2) We extract the 
 transition form factors by performing an overconstrained analysis of the
 lattice measurements  using all
lattice momentum vectors contributing to a given $Q^2$ value ~\cite{Negele}.

\vspace*{-0.3cm}

\section{Evaluation of the three-point function}
\begin{figure}[h]
\vspace*{-1cm}
{\mbox{\includegraphics[height=4.5cm,width=7cm]{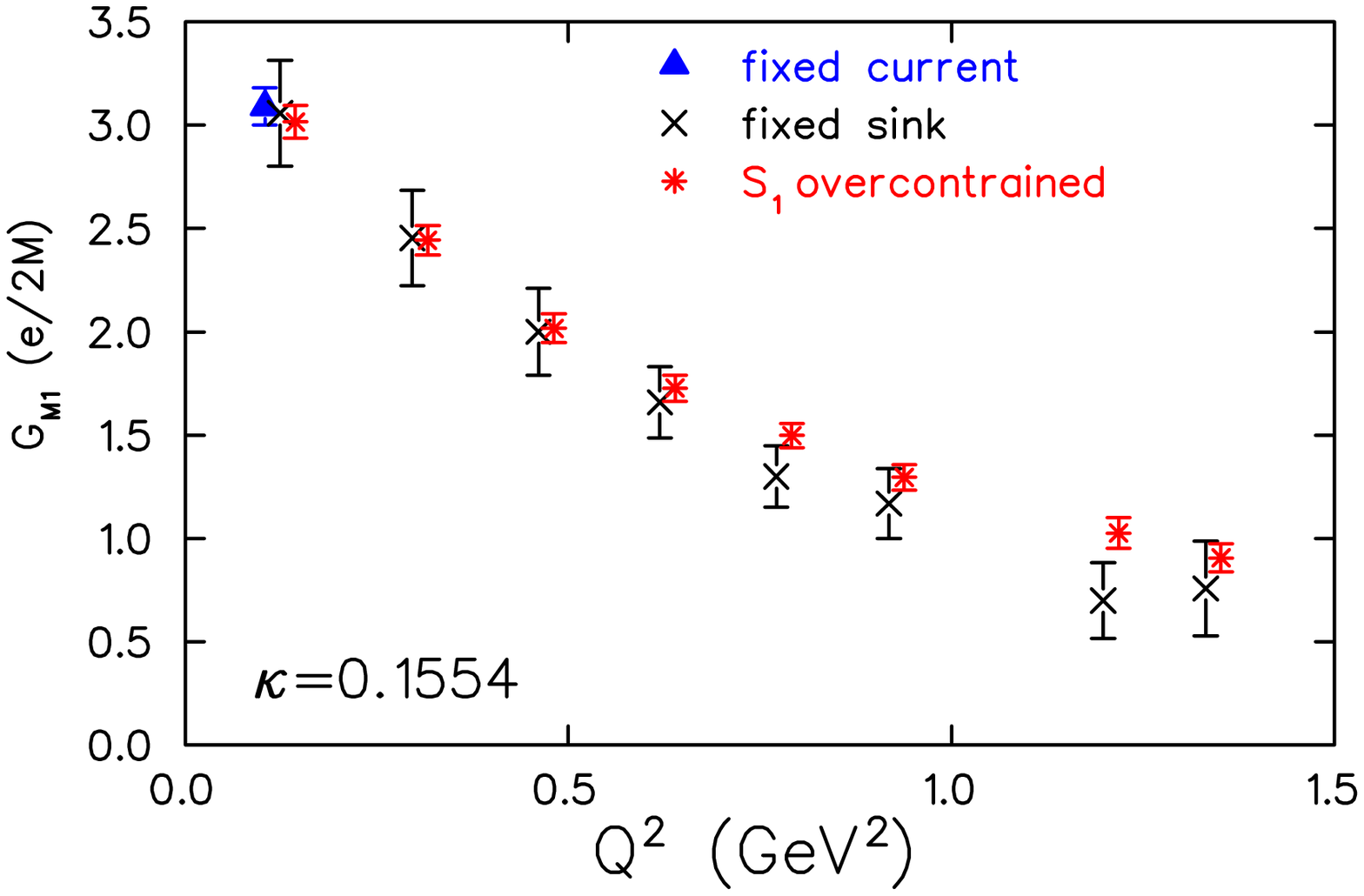}}}
\vspace*{-1cm}
\caption{ ${\cal G}_{M1}$ in Bohr magnetons
at $\kappa=0.1554$.
}
\vspace*{-0.8cm}
\label{fig:overconstrained}
\end{figure}

The evaluation  of 
the three-point function
$ G^{\Delta j^\mu N}_{\sigma} (t_2, t_1 ; {\bf p}^{\;\prime}, {\bf p};\Gamma )  $ 
can be done either using 
the  fixed current approach  as in
 previous lattice calculations~\cite{Leinweber,paper1}
or the fixed sink approach,
where the current  
can  couple to the backward sequential
propagator at any time slice $t_1$ carrying any lattice momentum~\cite{cairns},
 allowing the evaluation of the form factors at all possible momentum transfers.
As shown in
 Fig.~\ref{fig:overconstrained}, 
 for the same statistics, 
the errors in M1 in the fixed 
sink method are almost three times as large as
  those obtained in the fixed current approach.
Therefore 
in order to make use of the 
 advantages of the fixed sink approach, we must first reduce the errors.
We start by modifying the ratio
used in ref.~\cite{paper1} so that we do not need to evaluate both
$ \langle G^{\Delta j^\mu N}_{\sigma} (t_2, t_1 ; {\bf p}^{\;\prime}, {\bf p};\Gamma ) \rangle $ and 
 $ \langle G^{N j^\mu \Delta}_{\sigma} (t_2, t_1 ; {\bf p}^{\;\prime}, {\bf p};\Gamma ) \rangle $, since this would require two inversions.
Instead we use the ratio 
\small
\beq
\vspace*{-0.5cm}
&\>& \hspace*{-0.8cm}
R_\sigma=\frac{\langle G^{\Delta j^\mu N}_{\sigma} (t_2, t_1 ; {\bf p}^{\;\prime}, {\bf p};\Gamma ) \rangle \;}
{\langle G^{\Delta \Delta}_{ii} (t_2, {\bf p}^{\;\prime};\Gamma_4 ) \rangle \;} 
\biggr [\frac{\langle 
G^{\Delta \Delta}_{ii} (t_2, {\bf p}^{\;\prime};\Gamma_4 ) \rangle}{ \langle 
G^{N N} (t_2, {\bf p};\Gamma_4 ) \rangle }\> \nonumber \\
&\>& \hspace*{0.2cm}\frac{ \langle G^{N N}(t_2-t_1, {\bf p};\Gamma_4 ) \rangle \;\langle 
G^{\Delta \Delta}_{ii} (t_1, {\bf p}^{\;\prime};\Gamma_4 ) \rangle}
{\langle G^{\Delta \Delta}_{ii} (t_2-t_1, {\bf p}^{\;\prime};\Gamma_4 ) \rangle \;\langle 
G^{N N} (t_1, {\bf p};\Gamma_4 ) \rangle} \biggr ]^{1/2} \nonumber \\
&\;&\hspace*{1.2cm}\stackrel{t_2 -t_1 \gg 1, t_1 \gg 1}{\Rightarrow}
\Pi_{\sigma}({\bf p}^{\; \prime}, {\bf p}\; ; \Gamma ; \mu) \; ,
\label{R-ratio}
\eeq 
\normalsize
in the notation of ref.~\cite{paper1}.
We use kinematics where the $\Delta$ is produced at rest
and so ${\bf q}\equiv {\bf p}^{\prime}-{\bf p}=-{\bf p}$.
We fix $t_2/a=12$  and search for a plateau of 
 $ R_{\sigma}(t_2,t_1;{\bf p}^{\; \prime}, {\bf p}\; ; \Gamma ;\mu)$ as function of $t_1$.  

\normalsize

In the fixed sink approach, 
 the index $\sigma$ of the $\Delta$ 
and projection matrix $\Gamma$ are fixed 
 and therefore we need to determine the
most suitable choice of three-point functions from which to extract
the Sachs form factors ${\cal G}_{M1}, {\cal G}_{E2}$ and 
${\cal G}_{C2}$~\cite{paper1}. 
For example 
${\cal G}_{M1}$ can be extracted from
\be 
\Pi_{\sigma}({\bf q}\; ; \Gamma_4 ;\mu)= i A \epsilon^{\sigma 4\mu j} p^j {\cal G}_{M1}(Q^2)
\label{pure GM1}
\ee
where $A$ is a kinematical coefficient
and $\Gamma=\Gamma_4 = \frac{1}{2}
\left(\begin{array}{cc} I & 0 \\ 0 & 0 \end{array}
\right)$ in Eq.~(\ref{R-ratio}).
This leaves 3 choices for $\sigma$
 i.e there are three statistically independent  matrix elements
yielding ${\cal G}_{M1}$,
each requiring a sequential inversion. 
However, due to the $\epsilon$ factor,
fixing $\sigma$ 
means
that only momentum transfers in the other two directions contribute. Instead,
if we take the symmetric combination,
$S_1({\bf q};\mu)=\sum_{\sigma=1}^3
  \Pi_\sigma({\bf q}; \Gamma_4 ;\mu),$ 
momentum vectors in all directions contribute. This combination, which
we refer to as sink $S_1$, is built into the $\Delta$ interpolating
field and  requires only one inversion.
To take full advantage of the number of lattice vectors contributing
to  a given $Q^2$
we perform an overconstrained fit  by solving the overcomplete set of equations
$
P({\bf q};\mu)= D({\bf q};\mu)\cdot F(Q^2) 
$
where $P({\bf q};\mu)$ are the lattice measurements of the ratio
of Eq.~(\ref{R-ratio}),
$F =  \left(\begin{array}{c} {\cal G}_{M1} \\
                                   {\cal G}_{E2} \\ 
                                   {\cal G}_{C2} \end{array}\right)$
and, with $N$ being the number of current 
directions and momentum vectors contributing to 
a given $Q^2$,  D is an $N\times 3$ matrix which depends on 
kinematical factors. We extract the form factors by 
minimizing 
$
\chi^2=\sum_{k=1}^{N} \frac{1}{w_k^2}\left(\sum_{j=1}^3 D_{kj}F_j-P_k\right)^2,
$
where $w_k$ are the errors in the lattice
measurements, using singular value decomposition of D.
In Fig.~\ref{fig:overconstrained}, we compare the results for ${\cal G}_{M1}$
using an overconstrained analysis with sink type $S_1$ 
to our old analysis with  fixed 
$\sigma=2$ and $\mu=3$. 
The errors with our new analysis are reduced  at all
values of $Q^2$  and are now equal to the error obtained using the
fixed current approach.

\begin{figure}[t]
\vspace*{-1cm}
{\mbox{\includegraphics[height=9cm,width=7cm]{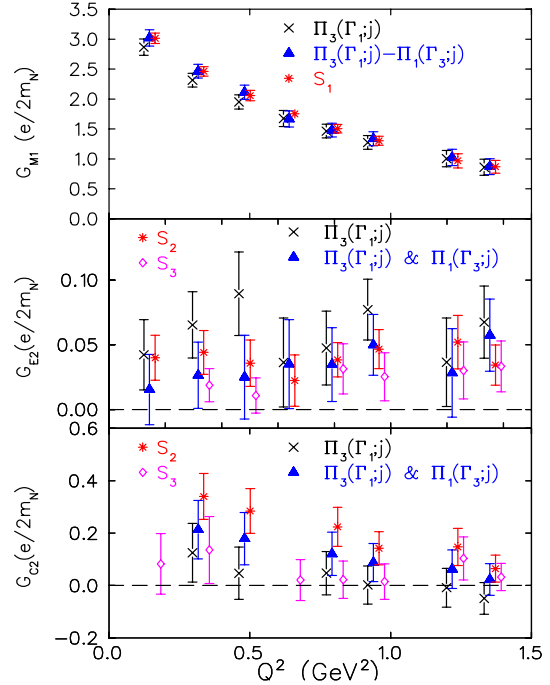}}}
\vspace*{-1cm}
\caption{ Top: ${\cal G}_{M1}$, middle: ${\cal G}_{E2}$,
bottom: ${\cal G}_{C2}$ in Bohr magnetons
at $\kappa=0.1554$ using $\Pi_3({\bf q};\Gamma_1;j)$ (crosses),
$\Pi_3({\bf q};\Gamma_1;j)$ \&  
$\Pi_1({\bf q};\Gamma_3;j)$ (filled triangles),
$S_1$ or $S_2$ (asterisks) and $S_3$ (open circles)  and an
overconstrained analysis.}
\vspace*{-0.8cm}
\label{fig:G all types}
\end{figure}

Using $\Gamma_k = \frac{1}{2}
\left(\begin{array}{cc} \sigma_k & 0 \\ 0 & 0 \end{array}
\right)$ instead of $\Gamma_4$  gives another six
statistically independent 3-point functions from which ${\cal G}_{M1}$ can be extracted: 
\beq
\vspace*{-0.5cm}
\Pi_{\sigma}({\bf q}\; ; \Gamma_k ;j)&=&  B \Biggl\{\frac{1}{2}
\left( p_\sigma\delta_{kj}-p_k\delta_{\sigma j}\right) {\cal G}_{M1}(Q^2) \nonumber\\ 
&\>&\hspace*{-2.5cm}-\biggl[ \frac{3}{2}\left( p_\sigma\delta_{kj}+p_k\delta_{\sigma j} \right ) 
            - \frac{3 p_\sigma p_k p_j}{{\bf p}^2} \biggr]  {\cal G}_{E2}(Q^2) 
\nonumber \\
&\>&\hspace*{-2.5cm}-\frac{(E_N-m_\Delta)}{2 m_\Delta}\>p_j \> \biggl(\delta_{\sigma k}-\frac{3 p_\sigma p_k}{{\bf p}^2} \biggr)  {\cal G}_{C2}(Q^2) \Biggr\}
\label{all G}
\eeq
where $B$ and $C$
are kinematical coefficients. 
To isolate the benefits of  using $S_1$   we compare in 
Fig.~\ref{fig:G all types}  ${\cal G}_{M1}$ 
 obtained using  $S_1$ to the 
the ones obtained by fixing
 $\sigma=3$ and $\Gamma_1$ in Eq.~(\ref{all G})~\cite{paper1}.  All results
are now obtained with the overconstrained analysis using 50 configurations.
As can be seen, $S_1$
produces results with the smallest errors 
and it is  therefore
the optimal sink for ${\cal G}_{M1}$.

For the extraction of the  quadrupole moments,
 we consider the symmetric combination
$S_2({\bf q};\mu)= \sum_{\sigma\neq k=1}^3 \Pi_\sigma({\bf q}; \Gamma_k ;\mu)$
from which both ${\cal G}_{E2}$ and
${\cal G}_{C2}$ can be extracted when the current is in the spatial direction.
 When the current is in the time direction, $S_2$ provides a statistically independent way for 
evaluating ${\cal G}_{C2}$, at no extra
cost. 
 Another combination to extract the quadrupole form factors is 
$S_3= \Pi_3({\bf q}\; ; \Gamma_3 ;\mu)
-\frac{1}{2}\bigl(\Pi_1({\bf q}\; ; \Gamma_1 ;\mu)+\Pi_2({\bf q}\; ; \Gamma_2 ;\mu)\bigr), 
$
 which, unlike $S_2$, contributes at the lowest value of $Q^2$. 
As  can be seen in Fig.~\ref{fig:G all types},
$S_2$ produces  results with smaller errors
as compared to those using $\Pi_3(\Gamma_1;j)$ and $\Pi_1(\Gamma_3;j)$ 
for both E2 and C2.
Increasing the the statistics from 50  to 200 configurations brings agreement
between  $S_2$ and $S_3$ for ${\cal G}_{C2}$ as well. 


\vspace*{-0.3cm}

\section{Results and Conclusions}

\begin{figure}[h]
\vspace*{-0.3cm}
{\mbox{\includegraphics[height=9cm,width=7cm]{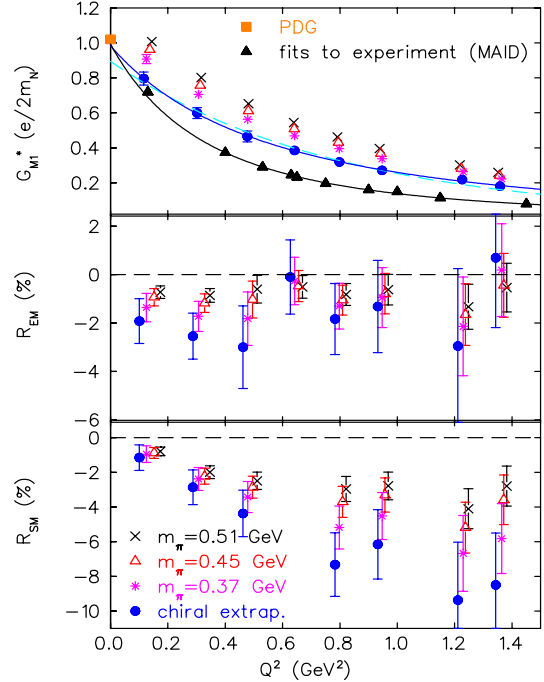}}}
\vspace*{-1cm}
\caption{Top: ${\cal G}_{M1}^*$, middle: EMR, bottom: CMR 
as function of $Q^2$ at $\kappa=0.1554$
(crosses), $\kappa=0.1558$ (open triangles), $\kappa=0.1562$ (asterisks)
and in the  chiral limit (filled circles). 
Filled triangles show  ${\cal G}_{M1}^*$ extracted
from measurements
 using MAID~\cite{Tiator}.
The dashed line is a fit to the lattice data using $a\exp(-bQ^2)$.}
\vspace*{-0.8cm}
\label{fig:G all}
\end{figure}

We analyse 200 configurations at 
 $\kappa= 0.1554,$ 0.1558 and 0.1562 
corresponding to   
$m_\pi/m_\rho=0.64,$ 0.59 and $0.50$ respectively. We use the nucleon
mass at the chiral limit to set the lattice spacing $a$, obtaining
$a^{-1}=2.04(2)$~GeV.
Using the optimal sink $S_1$  our results for ${\cal G}_{M1}^*$
are shown in Fig.~\ref{fig:G all}
as a function of $Q^2$,  where
\be 
\vspace*{-0.3cm}
{\cal G}_{M1}^* \equiv \frac{1}{3}\>\frac{1}{\sqrt{1+\frac{Q^2}{(m_N+m_\Delta)^2}}}
 \>{\cal G}_{M1} \quad.
\label{GM1*}
\vspace*{-0.3cm}
\ee
Results in the chiral
limit  are obtained by performing a linear extrapolation in  
 $m_\pi^2$. 
On the same figure, we also show the 
experimental values as extracted from the measured cross sections using the 
phenomenological model MAID~\cite{Tiator}. Although the lattice data in 
the chiral limit lie higher than the MAID data
both data sets are well described by the 
the phenomenological parametrization
$
{\cal G}_{a}(Q^2) = {\cal G}_a(0) \left(1 + \alpha Q^2 \right)\exp(-\gamma Q^2) G_E^p(Q^2)
$
for $a=M1, E2$ and $C2$ and 
$ G_E^p(Q^2)=1/(1+Q^2/0.71)^2$ is the proton electric form factor. 
These fits are shown in Fig.~\ref{fig:G all} by the solid lines.


In  Fig.~\ref{fig:G all},  we show  the ratios ${\rm EMR} \equiv R_{EM}= -\frac{{\cal G}_{E2}(q^2)}{{\cal G}_{M1}(q^2)}$
and ${\rm CMR} \equiv R_{SM}=-\frac{|{\bf q}|}{2m_\Delta}\;\frac{{\cal G}_{C2}(q^2)}{{\cal G}_{M1}(q^2)}$.
Results in the chiral limit are obtained  
by performing a  linear extrapolation in $m_\pi^2$. As expected, 
both EMR and
CMR become more negative as we approach the chiral limit. 
The results for EMR are  in agreement with experimental measurements whereas
CMR is not as negative as experiment at low $Q^2$. We believe that CMR is particularly sensitive to 
the absence of sea quarks and thus 
a  good probe of unquenching effects.

\end{document}